\documentstyle[12pt]{article}

\begin{document}

\title{
\begin{flushright}
\begin{small}
UPR/777-T \\
November 1997 \\
\end{small}
\end{flushright}
\vspace{1.cm}
Smarr's Formula for Black Holes in String Theory}

\author{Emre Sermutlu \\
\small Department of Physics and Astronomy\\
\small University of Pennsylvania\\
\small Philadelphia, PA 19104\\
\small e-mail: sermutlu@cvetic.hep.upenn.edu \\
\small (On Leave From) \\
\small Department of Mathematics \\
\small Bilkent University\\
\small Ankara, Turkey
\small e-mail:sermutlu@fen.bilkent.edu.tr
}
\date{}
\maketitle

\begin{abstract}
We investigate  thermodynamical properties of four and five dimensional
black hole solutions of toroidally compactified string theory. We derive 
an analog of Smarr's formula, and verify it directly using the
metric.
\end{abstract}

\section{Introduction}

Smarr's formula \cite{smr} has an important place in studies of black hole
thermodynamics. It gives us the variation of mass in terms of the variations in
entropy, angular momentum and electromagnetic charges, and provides us with an 
analogous thermodynamic relationship for the black hole system.

If we have a black hole solution, we can easily calculate the surface area of
the outer horizon
using the metric components. Entropy is given as $S={1\over 4 G_N}\, A$. Then,
if the
algebraic equations are tractable, we
can isolate $M$, take derivatives, and obtain Smarr's formula.

This procedure won't work if the solution is given in terms of parameters
that can't be solved explicitly in terms of mass and charges. Then we have to
use a roundabout way, using infinitesimal variations to make a change of
variables, which, in general involve inverting a big matrix, and if
all entries are nonzero, the results may be too complicated.

We expect the variation of mass with respect to area to be the temperature 
of the black hole, and
the variation of mass with respect to angular momentum to be the angular
velocity. These quantities can be computed from the metric. This will be an 
independent way to calculate the coefficients in the
Smarr's formula, and the results can be checked.

In this paper, we will follow the summarized procedure for two different
types of black
holes corresponding to four and five dimensional solutions of toroidally
compactified string theory. We first write the
area in terms of solution parameters, take the
infinitesimal variation of the area, and replace the solution parameters
by the physical ones using the jacobian matrix. Then we calculate $\Omega$
and $\kappa$ using the metric, and compare the results with the Smarr's
formula.

 In Section II, a four dimensional rotating black hole parametrized
by ADM mass, four charges, and angular momentum \cite{147}, and in Section III
a five
dimensional black hole with two angular momenta and three charges will be
analyzed. \cite{100}

Some of the related results for black holes in string theory have  been given
in \cite{ing},\cite{fin1},\cite{fin2}. A novel approach can be found in
\cite{kol}.

\section{Four Dimensions}
\subsection{The Metric and Physical Parameters}

The metric for four-dimensional rotating charged black hole solutions of
toroidally compactified superstring theory, 
parametrized by the ADM mass, four charges and angular
momentum, is given by \cite{147}

\begin{eqnarray}
ds^2_{E}&=&\Delta^{1\over 2}[-{{r^2-2mr+l^2{\cos}^2\theta}\over 
\Delta}dt^2+{{dr^2}\over{r^2-2mr+l^2}} + d\theta^2 
+{{{\sin}^2\theta}\over \Delta}\{(r+2m{\sinh}^2 
\delta_{1} )\cr 
&\times&(r+2m{\sinh}^2 \delta_{2}  )(r+2m{\sinh}^2 \delta_{3} )
(r+2m{\sinh}^2 \delta_{4}  )+l^2(1+{\cos}^2\theta)r^2+W \cr 
&+&2ml^2r{\sin}^2\theta\}d\phi^2
-{{4ml}\over \Delta}\{({\cosh} \delta_{1} {\cosh}
\delta_{2}  {\cosh} \delta_{3} {\cosh} \delta_{4}   \cr 
&-&{\sinh} \delta_{1} {\sinh} \delta_{2}  
{\sinh} \delta_{3} {\sinh} \delta_{4}  )r
+2m{\sinh}\delta_{1} {\sinh}\delta_{2}  {\sinh}\delta_{3} 
{\sinh}\delta_{4}  \}{\sin}^2 \theta dt d\phi],
\label{mt4}
\end{eqnarray}
where
\begin{eqnarray}
\Delta &\equiv& (r+2m{\sinh}^2 \delta_{1} )
(r+2m{\sinh}^2 \delta_{2}  )(r+2m{\sinh}^2 \delta_{3} )
(r+2m{\sinh}^2 \delta_{4}  )\cr
&+&(2l^2r^2+W){\cos}^2\theta, \cr
W &\equiv& 2ml^2({\sinh}^2\delta_{1} +{\sinh}^2\delta_{2}  +
{\sinh}^2\delta_{3} +{\sinh}^2\delta_{4}  )r \cr
&+&4m^2l^2(2{\cosh}\delta_{1} {\cosh}\delta_{2}  {\cosh}
\delta_{3} {\cosh}\delta_{4}  \,{\sinh}\delta_{1} {\sinh}
\delta_{2}  {\sinh}\delta_{3} {\sinh}\delta_{4}  \cr 
&-&2{\sinh}^2 \delta_{1} {\sinh}^2 \delta_{2}  {\sinh}^2 
\delta_{3} {\sinh}^2 \delta_{4}  -{\sinh}^2 \delta_{1}  
{\sinh}^2 \delta_{2} {\sinh}^2 \delta_{4}   \cr
&-&{\sinh}^2 \delta_{1} {\sinh}^2 \delta_{2} 
{\sinh}^2 \delta_{3}  -{\sinh}^2 \delta_{2} {\sinh}^2 
\delta_{3}  {\sinh}^2 \delta_{4}  -{\sinh}^2 \delta_{1} 
{\sinh}^2 \delta_{3}  {\sinh}^2 \delta_{4} )\cr
&+&l^4{\cos}^2 \theta.
\label{4ddef}
\end{eqnarray}

\noindent
The outer and inner event horizons are at

\begin{equation}
r_{\pm}=m\pm \sqrt{m^2-l^2},
\end{equation}

\noindent
Here, $m$, the non-extremality parameter, is related to the mass of Kerr
solution, $l$ is related to the
angular momentum of the Kerr solution and $\delta_{1,2,3,4}$ are 
boost parameters. Our
aim is to write the variation of $S$ in terms of the physical
parameters ADM mass $M$, two electric charges $Q_1, \,Q_2$, two magnetic
charges $P_1,\,P_2$, and the angular momentum $J$.  The physical parameters can
be expressed in terms of $m, l$ and the boosts as follows:

\begin{eqnarray}
M&=&4m(\cosh^2 \delta_{1} +\cosh^2 \delta_{2}  +\cosh^2 \delta_{3}  
+\cosh^2 \delta_{4} )-8m , \nonumber \\
Q_1&=&4m\cosh \delta_{1}  \,\sinh \delta_{1}  , \nonumber \\
Q_2 &=&4m\cosh \delta_{2}  \,\sinh \delta_{2}  ,  \\
P_1 &=&4m\cosh \delta_{3}  \,\sinh \delta_{3}  , \nonumber \\
P_2 &=&4m\cosh \delta_{4} \, \sinh \delta_{4}  , \nonumber \\
J &=&8\,l\,m\,(\cosh \delta_{1}  \cosh \delta_{2}  \cosh \delta_{3}  
\cosh \delta_{4} -\sinh \delta_{1} 
\sinh \delta_{2}  \sinh \delta_{3}  \sinh \delta_{4} ) .\nonumber 
\end{eqnarray}

\noindent
Note that we choose $G_N={\pi \over 4}$. 

\subsection{Smarr's Formula}

The entropy is given by $ {1\over 4 G_N}\,A$ where $A$ is the area of the outer
horizon. In this case, $S$ has the form: \cite{147}

\begin{eqnarray}
S&=&16 \pi [(m^2+m \sqrt{m^2-l^2})\, (\cosh \delta_{1}  \cosh \delta_{2} 
\cosh
\delta_{3}  \cosh \delta_{4} ) \nonumber \\
 & &+(m^2-m\sqrt{m^2-l^2})\, (\sinh \delta_{1}  \sinh \delta_{2}  
\sinh \delta_{3}  \sinh
\delta_{4} )].
\end{eqnarray}

\noindent
Note that we can write $S$ in the form 

\begin{equation}
S=16 \pi (m^2 K+m\sqrt{m^2-l^2} L),
\end{equation}

\noindent
where

\begin{eqnarray}
K&=&\cosh \delta_{1}  \, \cosh \delta_{2}  \, \cosh \delta_{3}  \, 
\cosh \delta_{4} +\sinh \delta_{1}  \,
\sinh \delta_{2}  \, \sinh \delta_{3}  \, \sinh \delta_{4}  ,\nonumber \\
L&=&\cosh \delta_{1}  \, \cosh \delta_{2}  \, \cosh \delta_{3}  \, 
\cosh \delta_{4} -\sinh \delta_{1}  \,
\sinh \delta_{2}  \, \sinh \delta_{3}  \, \sinh \delta_{4}  . 
\end{eqnarray}

\noindent
We can write the variation of entropy in terms of the solution parameters
as
follows:

\begin{equation}
d{S}={\partial S \over \partial \delta_1}\, d\delta_{1} +{\partial S \over 
\partial \delta_2}\, d\delta_{2} +{\partial S \over \partial \delta_3}\, 
d\delta_{3} +{\partial S \over \partial \delta_4}\, \delta_{4} 
+{\partial S \over \partial m}\, dm+{\partial S \over \partial l}\, dl,
\end{equation}

\noindent
But we want to write the variation in terms of the physical parameters:

\begin{equation}
dS=\Gamma_1\, dQ_1+\Gamma_2\, dQ_2+\Gamma_3\, dP_1+\Gamma_4
\,dP_4+\Gamma_5 \,dM+\Gamma_6\, dJ \label{gum},
\end{equation}

\noindent
where

\begin{equation}
\Gamma_1=\left( {\partial S \over \partial \delta_1}\, 
{\partial \delta_1 \over \partial Q_1}\,+{\partial S \over \partial \delta_2}\, 
{\partial \delta_2 \over \partial Q_1}\, +{\partial S \over \partial \delta_3}\, 
{\partial \delta_3 \over \partial Q_1}\, +{\partial S \over \partial \delta_4}\, 
{\partial \delta_4 \over \partial Q_1}\,+{\partial S \over \partial m}\, 
{\partial m \over \partial Q_1}\, +{\partial S \over \partial l}\, 
{\partial l \over \partial Q_1}\, \right),
\label{gam}
\end{equation}
\noindent
etc.

\noindent
By rearranging (\ref{gum}), we can now write the variation of M to obtain
an analog of Smarr's
formula:

\begin{equation}
dM=TdS+\Omega dJ+\Phi_1\,dQ_1+\Phi_2\,dQ_2+\Phi_3\,dP_1+\Phi_4\,dP_2,
\label{aci}
\end{equation}

\noindent
where

\begin{eqnarray}
T&=&{1\over \Gamma_5}, \nonumber \\
\Omega&=&-{\Gamma_6\over \Gamma_5}, \\
\Phi_i&=&-{\Gamma_i\over \Gamma_5} \;\;\;(i=1\ldots 4). \nonumber
\end{eqnarray}

\noindent
To determine the coefficients $\Gamma_i$, we have to invert the following
matrix:

\begin{equation}
\left( \begin{array}{l} dQ_1 \\ dQ_2 \\ dP_1 \\ dP_2 \\ dM \\ dJ \end{array} \right)
=\left( \begin{array}{cccccc}
4 m w_1 & 0 & 0 & 0 &  2 z_1 w_1 & 0\\
0 & 4 m w_2 & 0 & 0 & 2 z_2 w_2 & 0 \\
0 & 0 & 4 m w_3 & 0 & 2 z_3 w_3 & 0 \\
0 & 0 & 0 & 4 m w_4 & 2 z_4 w_4 & 0 \\
4 m z_1 w_1 & 4 m z_2 w_2 & 4 m z_3 w_3 & 4 m z_4 w_4& \hat{M} & 0 \\
8lmL_{1} & 8lmL_{2} & 8lmL_{3} & 8lmL_{4} & 8lL & 8mL
\end{array} \right)
\left(\begin{array}{l} d\delta_{1}  \\ d \delta_{2}  
\\ d \delta_{3}  \\ d \delta_{4}  \\ dm \\ dl \end{array} \right)
\end{equation}

\noindent
where

\begin{eqnarray}
w_i&=&\cosh 2 \delta_i, \nonumber \\
z_i&=&\tanh 2 \delta_i,  \\
\hat{M}&=&{M\over m}, \nonumber \\
L_i&=&{\partial L \over \partial \delta_i} \;\;\; (i=1, \ldots ,4). \nonumber
\end{eqnarray}

\noindent
The result is

\begin{equation}
\left(\begin{array}{l} d\delta_{1}  \\ d \delta_{2}  
\\ d \delta_{3}  \\ d \delta_{4}  \\ dm \\ dl \end{array} \right)
={1\over B} \left( \begin{array}{cccccc}
{B\over 4 m w_1} +z_1^2 & z_1 z_2 & z_1 z_3 & z_1 z_4 & -z_1 & 0 \\
z_2 z_1 &{B\over 4 m w_2}+  z_2^2 & z_2 z_3 & z_2 z_4 & -z_2 & 0 \\
z_3 z_1 & z_3 z_2 &  {B\over 4 m w_3}+z_3^2 & z_3 z_4 & -z_3 & 0 \\
z_4 z_1 & z_4 z_2 & z_4 z_3 &{B\over 4 m w_4}+  z_4^2 & -z_4 & 0 \\
-2mz_1 & -2mz_2 & -2mz_3 & -2mz_4 & 2m & 0 \\
{u_1\over L} & {u_2 \over L} & {u_3 \over L} & {u_4 \over L} &{lP_L\over L}-2 l& 
{B\over 8 m L} \\
\end{array} \right)
\left( \begin{array}{l} dQ_1 \\ dQ_2 \\ dP_1 \\ dP_2 \\ dM \\ dJ \end{array} \right)
\end{equation}

\noindent
where

\begin{eqnarray}
u_i&=&-z_i l P_L+2 z_i l L-{l L_i B\over 4 m w_i } \;\; (i=1,\ldots ,4), \nonumber
\\ 
P_L&=&L_{1} z_1+L_{2} z_2+L_{3} z_3+L_{4} z_4 , \\
B&=&{2 M}-4 m (w_1 z_1^2+w_2 z_2^2+w_3 z_3^2+w_4 z_4^2) .\nonumber 
\end{eqnarray}

\noindent
Now, using this matrix, we can calculate the $\Gamma_i$'s  defined
in (\ref{gam}).

\begin{eqnarray}
\Gamma_5&=&{S\over 4 \sqrt{m^2-l^2}} ,\nonumber \\
\Gamma_6&=&{-2 \pi l \over   \sqrt{m^2-l^2}}, \\
\Gamma_i&=&{-S_{i}\over 4 \sqrt{m^2-l^2}} \;\;(i=1,...,4), \nonumber
\end{eqnarray}

\noindent
where $S_i \equiv {\partial S \over \partial \delta_i}$. Thus, we can
find the coefficients in Smarr's formula (\ref{aci}) as: \footnote{Related to
the
formula obtained in \cite{fin2}}

\begin{eqnarray}
T&=&{4 \sqrt{m^2-l^2} \over S}, \nonumber \\
\Omega&=&{8 \pi l\over \,S}, \\
\Phi_{i}&=&{S_i \over S}. \nonumber
\end{eqnarray}

\subsection{Thermodynamical Quantities Derived From the Metric}

\noindent
Now, we determine the thermodynamical quantities entering (\ref{aci})
using the metric. The temperature $T$ is related to the surface gravity
$\kappa$ as

\begin{equation}
2\pi T=\kappa=-{1\over 2} {dg_{tt}\over dr} |_{r=r_+ , \theta=0}.
\end{equation}

\noindent
Using the metric

\begin{equation}
{dg_{tt}\over dr}|_{r=r_+ ,\theta=0} ={-2\,(r_+ -m)\over \sqrt{\Delta}},
\end{equation}

\noindent
where 

\begin{equation}
\sqrt{\Delta}|_{r=r_+ ,\theta=0}=2 m\, (m\, K+\sqrt{m^2-l^2}\, L)={S \over 8 \pi}.
\end{equation}

\noindent
Thus,

\begin{equation}
\kappa={8 \pi \sqrt{m^2-l^2} \over S},
\end{equation}

\noindent
which is in agreement with Smarr's formula (\ref{aci}).

\noindent
The angular velocity of the black hole at the outer horizon is:

\begin{equation}
\Omega \equiv {-g_{tt}\over g_{\phi t}}|_{r=r_+ ,\theta=0}. 
\end{equation}

\noindent
From the metric (\ref{mt4}) we can write

\begin{equation}
{g_{tt} \over g_{\phi t}}={r^2-2mr+l^2 -l^2 \sin^2 \theta 
\over 
2 m l \sin^2 \theta\, (r\,L+ m\, K -m\, L)}.
\end{equation}

\noindent
At the horizon, $r=r_+=m+\sqrt{m^2-l^2}$, which means $r^2+l^2-2mr=0$, so

\begin{equation}
{ g_{tt}\over g_{\phi t}}|_{r=r_+ ,\theta=0} 
={-l\over 2m (\sqrt{m^2-l^2} \, L+m\, K)}={-8 \pi l\over S},
\end{equation}

\begin{equation}
\Omega_H={8 \pi l \over S},
\end{equation}

which is also in agreement with Smarrs formula (\ref{aci}).

\section{ Black Holes in Five Dimensions}

\subsection{Metric and Physical Parameters}

The metric for five-dimensional rotating charged black holes of
toroidally compactified string theory, specified by the
ADM mass $M$, three charges $Q_1, Q_2, Q_3$ and two rotational parameters $l_1,
l_2$  is given by \cite{100}:

\begin{eqnarray}
ds_E^2&=&g_{tt}\,dt^2+g_{rr}\,dr^2+g_{\theta \theta} d\theta^2+2g_{\phi
\psi} d\phi d\psi+2g_{\phi t} d\phi dt \\
&&+2g_{\psi t} d\psi dt +g_{\phi\phi}d\phi^2+g_{\psi \psi} d\psi^2 , \nonumber
\end{eqnarray}

\begin{eqnarray}
g_{tt}&=&-\Delta^{-2\over 3} R (R-2m), \nonumber \\
g_{rr}&=&{\Delta^{1\over 3}\,r^2\over (r^2+l_1^2)(r^2+l_2^2)-2mr^2}, \nonumber \\
g_{\theta \theta}&=&\Delta^{1\over 3}, \nonumber \\
g_{\phi \psi}&=&\cos^2 \theta \sin^2 \theta \Delta^{-2\over 3} 
(L_1 k_3+L_3 k_1), \nonumber \\
g_{\phi t}&=&-2m \sin^2 \theta \Delta^{-2\over 3} 
(l_1 R c_1 c_2 c_3+l_2 (2m-R)s_1 s_2 s_3),  \\
g_{\psi t}&=&-2m \cos^2 \theta \Delta^{-2\over 3} 
(l_1 (2m-R)s_1 s_2 s_3+l_2 R c_1 c_2 c_3), \nonumber \\
g_{\phi \phi}&=&\sin^2 \theta \Delta^{-2\over 3} 
[\Delta+\sin^2 \theta(L_1 k_1+L_2 k_2+L_3 k_3)], \nonumber \\
g_{\psi \psi}&=&\cos^2 \theta \Delta^{-2\over 3} 
[\Delta+\cos^2 \theta(L_1 k_1-L_2 k_2+L_3 k_3)], \nonumber 
\end{eqnarray}

\noindent
where

\begin{eqnarray}
L_1&=&l_1^2+l_2^2, \nonumber \\
L_2&=&l_1^2-l_2^2, \nonumber \\
L_3&=&2 l_1 l_2  ,\nonumber \\
k_1&=&mR-2m^2 q-4m^2 t,\nonumber \\
k_2&=&R^2+mR+2mRp+2m^2q , \\
k_3&=&4m^2 \,c_1 c_2 c_3 s_1 s_2 s_3 , \nonumber \\
R&=&r^2+l_1^2 \cos^2 \theta +l_2^2 \sin^2 \theta ,\nonumber \\
p&=&s_1^2+s_2^2+s_3^2 , \nonumber \\
q&=&s_1^2 s_2^2+s_1^2 s_3^2+s_2^2 s_3^2 ,\nonumber \\
t&=&s_1^2 s_2^2 s_3^2 ,\nonumber
\end{eqnarray}

\noindent
and $s_i ,\; c_i$ stand for $\sinh \delta_i, \; \cosh \delta_i, \;
(i=1,2 , 3)$ respectively.
 
\noindent
Electromagnetic vector potentials are given as:

\begin{eqnarray}
A^{(1)}_{t\,1}&=&{{m{\cosh}\delta_{1} {\sinh}\delta_{1} }
\over{r^2+2m{\sinh}^2\delta_{1} +l^2_1{\cos}^2\theta +
l^2_2{\sin}^2 \theta}}, \cr
A^{(1)}_{\phi\,1}&=&m{\sin}^2\theta
{{l_1{\sinh}\delta_{1} {\sinh}\delta_{2} {\cosh}\delta_{3}
-l_2{\cosh}\delta_{1} {\cosh}\delta_{2} {\sinh}\delta_{3}}
\over{r^2+2m{\sinh}^2\delta_{1} +l^2_1{\cos}^2\theta +
l^2_2{\sin}^2 \theta}}, \cr
A^{(1)}_{\psi\,1}&=&m{\cos}^2\theta
{{l_1{\cosh}\delta_{1} {\sinh}\delta_{2} {\sinh}\delta_{3}
-l_2{\sinh}\delta_{1} {\cosh}\delta_{2} {\cosh}\delta_{3}}
\over{r^2+2m{\sinh}^2\delta_{1} +l^2_1{\cos}^2\theta +
l^2_2{\sin}^2 \theta}}, \cr
A^{(2)}_{t\,1}&=&{{m{\cosh}\delta_{2} {\sinh}\delta_{2} }
\over{r^2+2m{\sinh}^2\delta_{2} +l^2_1{\cos}^2\theta +
l^2_2{\sin}^2 \theta}}, \cr
A^{(2)}_{\phi\,1}&=&m{\sin}^2\theta
{{l_1{\cosh}\delta_{1} {\sinh}\delta_{2} {\cosh}\delta_{3}
-l_2{\sinh}\delta_{1} {\cosh}\delta_{2} {\sinh}\delta_{3}}
\over{r^2+2m{\sinh}^2\delta_{2} +l^2_1{\cos}^2\theta +
l^2_2{\sin}^2 \theta}}, \cr
A^{(2)}_{\psi\,1}&=&m{\cos}^2\theta
{{l_1{\sinh}\delta_{1} {\cosh}\delta_{2} {\sinh}\delta_{3}
-l_2{\cosh}\delta_{1} {\sinh}\delta_{2} {\cosh}\delta_{3}}
\over{r^2+2m{\sinh}^2\delta_{2} +l^2_1{\cos}^2\theta +
l^2_2{\sin}^2 \theta}}, \cr
B_{t\phi}&=&-2m\sin^2\theta(l_1\sinh\delta_{1} \sinh\delta_{2} \cosh
\delta_{3} -l_2\cosh\delta_{1} \cosh\delta_{2} \sinh\delta_{3} )(r^2+l^2_1
\cos^2\theta \cr
& &\ \ +l^2_2\sin^2\theta+m\sinh^2\delta_{1} +m\sinh^2\delta_{2} )/ 
[(r^2+l^2_1\cos^2\theta+l^2_2\sin^2\theta+2m\sinh^2\delta_{1} ) \cr 
& & \ \ \times(r^2+l^2_1\cos^2\theta+l^2_2\sin^2\theta +
2m\sinh^2\delta_{2} )], \cr
B_{t\psi}&=&-2m\cos^2\theta(l_2\sinh\delta_{1} \sinh\delta_{2} 
\cosh\delta_{3} -l_1\cosh\delta_{1} \cosh\delta_{2} \sinh\delta_{3} )
(r^2+l^2_1\cos^2\theta \cr
& & \ \ +l^2_2\sin^2\theta+m\sinh^2\delta_{1} +m\sinh^2\delta_{2} )/ 
[(r^2+l^2_1\cos^2\theta+l^2_2\sin^2\theta+2m\sinh^2\delta_{1} ) \cr 
& &\ \ \times(r^2+l^2_1\cos^2\theta+l^2_2\sin^2\theta+
2m\sinh^2\delta_{2} )], \cr
B_{\phi\psi}&=&{{2m\cosh\delta_{3} \sinh\delta_{3} \cos^2\theta\sin^2\theta
(r^2+l^2_1\cos^2\theta+l^2_2\sin^2\theta+m\sinh^2\delta_{1} +
m\sinh^2\delta_{2} )} \over {(r^2+l^2_1\cos^2\theta+l^2_2\sin^2\theta+
2m\sinh^2\delta_{1} )(r^2+l^2_1\cos^2\theta+l^2_2\sin^2\theta+
2m\sinh^2\delta_{2} )}}, \cr
\end{eqnarray}

\begin{equation}
\Delta=R^3+2mpR^2+4m^2qR+8m^3t
\end{equation}

\begin{equation}
r_{\pm}^2=m-{1\over 2} L_1\pm {1\over 2} \sqrt{L_2^2+4m(m-L_1)}
\end{equation}

\noindent
We choose $G_N={\pi \over 4}$.

\noindent
The physical quantities ADM mass $M$, three charges $Q_1,Q_2,Q_3$ and two
angular momenta $J_1,J_2$  are given as

\begin{eqnarray}
M&=&2m(\cosh^2 \delta_{1} +\cosh^2 \delta_{2}  +\cosh^2 \delta_{3}  )-3m , 
\nonumber \\
Q_1&=&2m\cosh \delta_{1}  \,\sinh \delta_{1}  ,  \nonumber \\
Q_2 &=&2m\cosh \delta_{2}  \,\sinh \delta_{2}  ,  \\
Q_3 &=&2m\cosh \delta_{3}  \,\sinh \delta_{3}  , \nonumber \\
J_1 &=&4\,m\,(l_1\,\cosh \delta_{1}  \cosh \delta_{2}  \cosh \delta_{3}  
-l_2\,\sinh \delta_{1} 
\sinh \delta_{2}  \sinh \delta_{3} ) , \nonumber \\
J_2 &=&4\,m\,(l_2\,\cosh \delta_{1}  \cosh \delta_{2}  \cosh \delta_{3}  
-l_1\,\sinh \delta_{1} 
\sinh \delta_{2}  \sinh \delta_{3} ), \nonumber 
\end{eqnarray}

\noindent
where $m$ is the nonextremality parameter, $\delta_{1,2,3}$ are the boost
parameters and $l_{1,2}$ are the angular momentum parameters.

\subsection{Smarr's Formula}

The entropy is given by:\cite{147}

\begin{eqnarray}
S&=&4 \pi m [\sqrt{2m-(l_1-l_2)^2}\, (\cosh \delta_{1}  \cosh \delta_{2} 
\cosh \delta_{3}  +\sinh \delta_{1}  \sinh \delta_{2}  
\sinh \delta_{3}  ) \nonumber \\
 & &+\sqrt{2m-(l_1+l_2)^2} (\cosh \delta_{1}  \cosh \delta_{2} 
\cosh \delta_{3}  -\sinh \delta_{1}  \sinh \delta_{2}  \sinh \delta_{3}  )]
\end{eqnarray}

\begin{equation}
dS=\Gamma_1\, dQ_1+\Gamma_2\, dQ_2+\Gamma_3\, dQ_3+\Gamma_4
\,dM+\Gamma_5 \,dJ_1+\Gamma_6\, dJ_2 \label{sss}.
\end{equation}

\noindent
To find the derivatives of boosts with respect to physical variables
( ${\partial \delta_1 \over \partial Q_2}\,$ etc.) 
we need to invert the following matrix:

\begin{equation}
\left( \begin{array}{l} dQ_1 \\ dQ_2 \\ dQ_3  \\ dM 
\\ dJ_1 \\ dJ_2 \end{array} \right)
=\left( \begin{array}{cccccc}
2 m w_1 & 0 & 0 &  z_1 w_1 & 0 & 0\\
0 & 2 m w_2 & 0 &  z_2 w_2 & 0 & 0\\
0 & 0 & 2 m w_3 &  z_3 w_3 & 0 & 0\\
2 m z_1 w_1 & 2 m z_2 w_2 & 2 m z_3 w_3 & w_0+w_1+w_2& 0&0 \\
J_{1,1} & J_{1,2} & J_{1,3} & {J_1/ m} & 4mC &-4mE \\
J_{2,1} & J_{2,2} & J_{2,3} & {J_2/ m} & -4mE &4mC 
\end{array} \right)
\left(\begin{array}{l} d\delta_{1}  \\ d \delta_{2}  
\\ d \delta_{3}  \\ dm \\ dl_1 \\dl_2  \end{array} \right)
\end{equation}

\noindent
where

\begin{eqnarray}
w_i&=&\cosh 2 \delta_i \;\;(i=1,2,3) ,\nonumber \\
z_i&=&\tanh 2 \delta_i \;\;(i=1,2,3) , \nonumber \\
C&=&\cosh \delta_{1}  \, \cosh \delta_{2}  \, \cosh \delta_{3}  , \nonumber \\
E&=&\sinh \delta_{1}  \,\sinh \delta_{2}  \, \sinh \delta_{3}  .\nonumber \\
\end{eqnarray}

\noindent
The result is

\begin{equation}
\left(\begin{array}{l} d\delta_{1}  \\ d \delta_{2}  
\\ d \delta_{3}  \\ dm \\ dl_1 \\dl_2 \end{array} \right)
={1\over 2 m\, U } \left( \begin{array}{cccccc}
{ U\over w_1} + z_1^2 &  z_1 z_2 &  z_1 z_3  & - z_1 & 0 & 0 \\
 z_2 z_1 &{ U\over w_2}+ z_2^2 &  z_2 z_3 & - z_2 & 0 & 0 \\
 z_3 z_1 &  z_3 z_2 &  { U\over  w_3}+ z_3^2  & - z_3 & 0 &0 \\
-2m  z_1 & -2m  z_2 & -2m  z_3  & 2m  & 0 &0 \\
t_{51} & t_{52} & t_{53} & -c & {C U/ (2 h)} & {E U /(2h)} \\
t_{61} & t_{62} & t_{63} & -d & {E U/ (2 h)} & {C U /(2h)} \\
\end{array} \right)
\left( \begin{array}{l} dQ_1 \\ dQ_2 \\ dQ_3 \\  dM \\ dJ_1 
\\dJ_2 \end{array} \right)
\end{equation}

\noindent
where

\begin{eqnarray}
U&=&(w_1+w_2+w_3-w_1 z_1^2-w_2 z_2^2-w_3 z_3^2), \nonumber \\
P_E&=&E_1 z_1+E_2 z_2+E_3 z_3 , \nonumber \\
P_C&=&C_1 z_1+C_2 z_2+C_3 z_3  , \nonumber \\
h&=&(C-E)(C+E) , \nonumber \\
a&=&{E\,l_1+C\,l_2\over h} ,  \nonumber \\
b&=&{C\,l_1+E\,l_2\over h} , \\
c&=&2  l_1+a\,P_E-b\,P_C  ,\nonumber \\
d&=&2  l_2+b\,P_E-a\,P_C , \nonumber \\ 
t_{5i}&=&{U \over w_i}(E_i \, a-C_i \, b)+z_i c , \nonumber \\
t_{6i}&=&{U \over w_i}(E_i \, b-C_i \, a)+z_i d , \nonumber
\end{eqnarray}

\noindent
Using these results, we can calculate $\Gamma_i 's$ as follows:

\begin{eqnarray}
\Gamma_i&=&{-S_i\over \alpha \beta}, \nonumber \\
\Gamma_4&=&{S\over \alpha \beta } , \\
\Gamma_5&=&\pi  \left({l_2-l_1\over \alpha}-{l_1+l_2\over \beta} \right) , \\
\Gamma_6&=&\pi  \left({l_1-l_2\over \alpha}
-{l_1+l_2\over \beta} \right) , \nonumber
\end{eqnarray}

\noindent
where

\begin{eqnarray}
\alpha&=&\sqrt{2m-(l_1-l_2)^2},\\
\beta&=&\sqrt{2m-(l_1+l_2)^2} .\nonumber
\end{eqnarray}

\noindent
The Smarr's formula is of the form:

\begin{equation}
dM=T\,dS+\Phi_1\,dQ_1 +\Phi_2 \,dQ_2+\Phi_3 \, dQ_3+\Omega_1
\,dJ_1+\Omega_2\,dJ_2 \label{sm2},
\end{equation}

\noindent
where

\begin{eqnarray}
T&=&{\alpha \beta \over  S}, \nonumber \\
\Phi_i&=&{ S_{,i}\over S} (i=1,2,3) ,\\
\Omega_1&=&-\pi {\beta(l_2-l_1)-\alpha(l_1+l_2) \over  S}, \nonumber \\
\Omega_2&=&-\pi {\beta (l_1-l_2)-\alpha(l_1+l_2)\over  S}. \nonumber \\
\end{eqnarray}

\subsection{Thermodynamic Quantities Derived from the Metric}

\noindent
Now, we make an independent
check for the coefficients in Smarr's formula. Using the metric, we can
calculate $\Omega_1$.

\begin{eqnarray}
\Omega_1& \equiv &{-g_{tt}\over g_{\phi t}}|_{r=r_+ ,\theta={\pi \over 2}},\\
 &=& {R (R-2m) \over -2m \sin^2 \theta  (l_1 R c_1 c_2 c_3+l_2 (2m-R)s_1s_2s_3)}.
\nonumber
\end{eqnarray}

\noindent
where $R=r^2+l_1^2 \cos^2 \theta+l_2^2 \sin^2 \theta$. At the
outer horizon,
$r=r_+,\;\;\theta={\pi \over 2}$.

\begin{equation}
R=-{[\beta(l_2-l_1)-\alpha(l_2+l_1)]\,l_2\over \alpha-\beta},\;\;\;
R-2m={[\beta(l_2-l_1)-\alpha(l_2+l_1)] l_1\over \alpha+\beta}
\end{equation}

\noindent
So

\begin{equation}
\Omega_1=-2 \pi {\beta(l_2-l_1)-\alpha(l_2+l_1)\over S}
\end{equation}

\noindent
This result is in agreement with Smarr's formula (\ref{sm2}) except for a 
numerical factor. We can repeat
the calculation for $\Omega_2$. It is also in agreement with Smarr's formula.
Now, let us make an independent check for $\kappa$, the surface gravity of the
outer horizon. \cite{wal}

\begin{equation}
2 \pi T=\kappa=-{1\over 2} \sqrt{-g^{rr} \, g^{tt}}\; {d\Sigma\over dr}|_{r=r_+,
\theta={\pi \over 2}},
\end{equation}

\noindent
where

\begin{equation}
\Sigma=g_{tt}-{(g_{\phi t}+g_{\psi t})^2\over g_{\phi \phi}+g_{\psi
\psi}+2g_{\phi \psi}}.
\end{equation}

\noindent
We know that $g^{rr}=g_{rr}^{-1}$ and $g^{tt}=(g_{\phi \phi} 
g_{\psi \psi}-g_{\phi \psi}^2)/D$, where

\begin{eqnarray}
D&=&{Det(g_{ij})\over g_{rr} g_{\theta \theta}} \\
D&=&g_{tt} (g_{\phi \phi} g_{\psi \psi}-g_{\phi t}^2)
+2 g_{\phi \psi} g_{\phi t} g_{\psi t}-g_{\psi t}^2 g_{\phi \phi}
-g_{\phi t}^2 g_{\psi \psi} \nonumber
\end{eqnarray}

\noindent
After some algebra, we find that

\begin{eqnarray}
D&=&\cos^2 \theta \sin^2 \theta [(2m-R)R+L_2 (\cos^2 \theta-\sin^2 \theta) (R-m)
-L_1 m+L_2^2 \cos^2 \theta \sin^2 \theta] \nonumber \\
&=& -\cos^2 \theta \sin^2 \theta \, \Delta^{1\over 3} \, r^2 \,g_{rr}^{-1} 
\end{eqnarray}

\begin{eqnarray}
g_{\phi \phi} g_{\psi \psi}-g_{\phi \psi}^2&
=&\cos^2 \theta \sin^2 \theta \Delta^{-1/3} [\Delta+k_1 L_1
+(\sin^2 \theta-\cos^2 \theta) k_2 L_2 \nonumber \\
&&+k_3 L_3-\cos^2 \theta \sin^2 \theta L_2^2 (2m+2mp+R)]
\end{eqnarray}

\noindent
We know that $g^{rr}=g_{rr}^{-1}$, so

\begin{equation}
-g^{tt}\,g^{rr}={\Delta+k_1 L_1+(\sin^2 \theta -\cos^2 \theta) k_2 L_2
+k_3 L_3-\cos^2 \theta \sin^2 \theta L_2^2
(2m+2mp+R) \over r^2\,\Delta^{2\over 3}}
\end{equation}

\begin{equation}
\Sigma=\Delta^{1\over 3} {R(2m-R)+R(\cos^2 \theta -\sin^2 \theta) L_2
-2m(\cos^2 \theta l_1+\sin^2 \theta l_2)^2 \over
\Delta+(L_1 k_1+L_3 k_3) (\sin^4 \theta+\cos^4 \theta)+L_2 k_2 (\sin^2 \theta
-\cos^2 \theta)+2\sin^2 \theta \cos^2 \theta (L_1 k_3+L_3 k_1)}
\end{equation}

Note that  at the horizon, $R=m-{1\over 2} L_2+{1\over 2} 
\sqrt{L_2^2+4 m(m-L_1)}$ and
$\Sigma=0$.

\noindent
We find
\begin{equation}
\kappa={\sqrt{l_2^2+4m(m-L_1)}\over \sqrt{\Delta+L_1 k_1+L_2 k_2+L_3 k_3}}
\end{equation}

\noindent
At the horizon, $S=4 \pi \sqrt{\Delta+L_1 k_1+L_2 k_2+L_3 k_3}$.

So, 

\begin{equation}
{\kappa}={4 \pi \alpha \beta \over S}
\end{equation}

We can check the potentials $\Phi_i$ for the special case $l_1=l_2=0$. In this
case, $r_+^2=2m$, and $\Phi_1=A_{t1}^{(1)}$, $\Phi_2=A_{t1}^{(2)}$.

\section{Conclusion}
In this paper, we have calculated Smarr's formula for two different black
holes, and then made an independent check using the metric coefficients.
Temperature and angular momentum found in two different ways are in
agreement. 

Note that in four dimensions, we used the formula $\kappa=-{1\over 2}
{dg_{tt} \over dr}$, but in five dimensions we have to use the more
general formula

\begin{equation}
\kappa=-{1\over 2} \sqrt{-g^{rr} \, g^{tt}}\; {d\over dr}
\left(g_{tt}-{(g_{\phi t}+g_{\psi t})^2\over g_{\phi \phi}+g_{\psi
\psi}+2g_{\phi \psi}}\right)|_{r=r_+,
\theta={\pi \over 2}},
\end{equation}

\newpage

{\Large \bf Acknowledgements:}
\newline

I would like to  thank Mirjam Cvetic for giving me the idea, expanding it
through stimulating discussions, for her guidance, helpful comments and
her interest  throughout the work. I would also like to thank Finn Larsen
and Metin Gurses
for helpful suggestions. This work was supported by the BDP program of 
TUBITAK (Scientific and Technical Research Council of Turkey).

\end{document}